\documentclass[12pt]{article}
\usepackage{graphicx}
\usepackage{xspace}

\newcommand\pubnumber{}
\newcommand\pubdate{\today}

\def\C{\ensuremath{C}\xspace}
\def\P{\ensuremath{P}\xspace}
\def\T{\ensuremath{T}\xspace}
\def\CP{\ensuremath{CP}\xspace}
\def\CPT{\ensuremath{CPT}\xspace}

\usepackage{relsize}
\def\babar{\mbox{\slshape B\kern-0.1em{\smaller A}\kern-0.1em
    B\kern-0.1em{\smaller A\kern-0.2em R}}\xspace}

\def\qmul{School of Physics and Astronomy\\
Queen Mary University of London, London, E1 4NS, UK}
\def\support{\footnote{Work supported by the Science and Technology Facilities Council, UK.}}

\def\Title#1{\begin{center} {\Large #1 } \end{center}}
\def\Author#1{\begin{center}{ \sc #1} \end{center}}
\def\Address#1{\begin{center}{ \it #1} \end{center}}

\newcommand\pubblock{\rightline{\begin{tabular}{l} \pubnumber\\
         \pubdate  \end{tabular}}}
\newenvironment{Abstract}{\begin{quotation}  }{\end{quotation}}
\newenvironment{Presented}{\begin{quotation} \begin{center} 
             PRESENTED AT\end{center}\bigskip 
      \begin{center}\begin{large}}{\end{large}\end{center} \end{quotation}}
\def\Acknowledgements{\bigskip  \bigskip \begin{center} \begin{large}
             \bf ACKNOWLEDGEMENTS \end{large}\end{center}}

\def\beq{\begin{equation}}
\def\eeq#1{\label{#1}\end{equation}}
\def\eeqn{\end{equation}}

\def\beqa{\begin{eqnarray}}
\def\eeqa#1{\label{#1}\end{eqnarray}}
\def\eeqan{\end{eqnarray}}

\let\bar=\overbar

\def\Dslash{\not{\hbox{\kern-4pt $D$}}}
\def\dslash{\not{\hbox{\kern-2pt $\del$}}}

\def\msb{{\bar{\ssstyle M \kern -1pt S}}}

\begin{document}
\begin{titlepage}
\pubblock

\vfill
\Title{$C$, $P$, and $CP$ asymmetry observables based on triple product asymmetries}
\vfill
\Author{Adrian J. Bevan\support}
\Address{\qmul}
\vfill
\begin{Abstract}
The discrete symmetries $C$, $P$ and $CP$ are known to be violated by the weak interaction.  It is possible
to probe the breaking of these symmetries using asymmetries constructed from triple products based on the 
decay of some particle $M$ to a four body final state.  These proceedings discuss the full set of possible
asymmetries that can be probed and applications to various measurement scenarios, focusing mostly on charm
mesons and baryons.  The ramifications of what can be learned from such measurements are also discussed.
\end{Abstract}
\vfill
\begin{Presented}
The 7th International Workshop on Charm Physics (CHARM 2015)\\
Detroit, MI, 18-22 May, 2015
\end{Presented}
\vfill
\end{titlepage}
\def\thefootnote{\fnsymbol{footnote}}
\setcounter{footnote}{0}
%

\section{Introduction}

The discrete symmetries of parity ($P$), charge conjugation ($C$), time (or motion) reversal ($T$),
and the combination $CP$ are known to be violated in weak interactions, but are conserved in 
both electromagnetic and strong interactions.  The overall combination $CPT$ is experimentally
conserved in locally gauge invariant quantum field theories such as the Standard Model of particle
physics (SM), and a corollary of this is that 
Lorentz symmetry is also conserved in such theories.  At some scale it is expected that $CPT$
will be violated as our understanding of subatomic physics adapts to account for quantum 
gravity.  Parity violation was discovered by Wu {et al.} in 1957~\cite{Wu:1957my},
and \CP was found to be violated in 1964 by Christenson {et al.}~\cite{Christenson:1964fg}.
For the past 50 years tests of $CP$ violation in quark interactions have provided a 
consistent picture: all known violations are consistent with the 1964 discovery.
Most discrete symmetry violation measurements have concentrated on testing \CP given the implications
noted by Sakharov that both \C and \CP violation are required for the Universe to evolve from 
the Big Bang to a matter dominated state as observed today~\cite{Sakharov:1967dj}.  However, the observed
level of \CP violation is insufficient to be able to explain this observed matter dominance of the
Universe.    These proceedings discuss tests of these symmetries using asymmetries obtained using
scalar triple products constructed from spins or momenta of the final state particles. 
We can consider the decay of some particle $M$ to a four body final
state $abcd$ and the $C$ conjugate process $\overline M\to \overline{abcd}$, where $ab$ and $cd$
can be used to construct decay planes in the centre of mass of the decaying particle\footnote{Charge 
conjugation is implied throughout unless otherwise specified, and the pairings defined depend on the 
possible physical states that are reconstructable.  A more general amplitude analysis based on a model
summing over all relevant interfering amplitudes is desirable, however construction of a relevant model
to fit to data requires experimental input.}.  Considering the three momenta
of the final state particles it is straight forward to see that a scalar triple product
$\vec{p}_c \cdot (\vec{p}_a \times \vec{p}_b)$ is even under $C$ and $CPT$ and odd under $P$, $T$ and $CP$.
To test a symmetry we need to identify a process $\psi \equiv M\to abcd$ and the conjugate under
the symmetry $S$ of interest $\psi^\prime = S \psi$.  From this conjugate pairing we are able
to construct an asymmetry given by the probabilities for $\psi$ and $\psi^\prime$ to occur:
\begin{eqnarray}
A = \frac{ P(\psi^\prime) - P(\psi)} {P(\psi^\prime) + P(\psi)}.\label{eq:asym}
\end{eqnarray}
A second asymmetry can be constructed by considering the \C conjugate decay under the symmetry.
If $A$ is non-zero these probabilities differ and the symmetry is violated.
We can write down the rate for decays of $M$ with a positive triple product (upward going decay)
as $\Gamma_+$ and that with a negative triple product (downward going) as $\Gamma_-$.  The
corresponding anti-particle rates are denoted by $\overline\Gamma_\pm$.  By considering 
\C, \P and \CP acting on the four $\Gamma$s we can construct six distinct asymmetries; 
these are~\cite{Bevan:2014nva}
\begin{eqnarray}
A_{\P} = \frac{\Gamma_{+} - \Gamma_{-}} {\Gamma_{+} + \Gamma_{-}}, &\,\,\,\,\,&
\overline{A}_{\P} = \frac{\overline\Gamma_{+} - \overline\Gamma_{-}} {\overline\Gamma_{+} + \overline\Gamma_{-}},\label{eq:tp:parity}\\
A_{\C} = \frac{\overline\Gamma_{-} - \Gamma_{-}} {\overline\Gamma_{-} + \Gamma_{-}}, &\,\,\,\,\,&
\overline A_{\C} = \frac{\overline\Gamma_{+} - \Gamma_{+}} {\overline\Gamma_{+} + \Gamma_{+}},\label{eq:tp:c}\\
A_{\CP} = \frac{\overline{\Gamma}_{+} - \Gamma_{-}} {\overline{\Gamma}_{+} + \Gamma_{-}}, &\,\,\,\,\,&
\overline{A}_{\CP} = \frac{\overline\Gamma_{-} - \Gamma_{+}} {\overline\Gamma_{-} + \Gamma_{+}}.\label{eq:tp:cp}
\end{eqnarray}
The subscript on $A$ denotes the symmetry used to construct the conjugate pair.    
We can construct another six asymmetries by considering the remaining symmetry transformations
on $A$ and $\overline A$ in turn. These are given by~\cite{Bevan:2014nva}
\begin{eqnarray}
a_{\C}^{\P}&=& \frac{1}{2}\left( A_{\P} - \overline{A}_{\P}\right), \nonumber\\
a_{\CP}^{\P} &=& \frac{1}{2}\left( A_{\P} + \overline{A}_{\P}\right),\nonumber\\
a_{\P}^{\C} &=& \frac{1}{2}(A_{\C} - \overline{A}_{\C}),\nonumber\\
a_{\CP}^{\C} &=& \frac{1}{2}(A_{\C} + \overline{A}_{\C}),\nonumber\\
a^{\CP}_{\P} &=& \frac{1}{2}(A_{\CP} - \overline A_{\CP}),\nonumber\\
a^{\CP}_{\C} &=& \frac{1}{2}(A_{\CP} + \overline A_{\CP}).\label{eq:tp:derived}
\end{eqnarray}
Here these secondary asymmetries are denoted by a lower case $a$, the superscript corresponds to the first symmetry
used and the subscript corresponds to the second one.  
We can determine which symmetry is under scrutiny by multiplying the superscripts and subscripts together.  
For example $a_{\C}^{\P}$ is a test of \CP, $a_{\CP}^{\P}$ a test of $\C\P^2 = \C$ etc.
In general these asymmetries receive contributions
from all interactions, and the interest is in isolating effects that are dominated by weak interactions (i.e. theoretically
clean), or where that may not be possible, to isolate effects that can either be understood in the longer term, or
signify a non-trivial weak interaction effect even in the presence of pollution from strong force
induced final state interactions (i.e. soft QCD and re-scattering).
A concrete example of this issue is discussed below in the context of the measurement of $\alpha_b$ from $\Lambda_b$ decays.
Hence interpretation of these asymmetries depends on the decay under study (and hence the model required to interpret data),
and an example where one considers two interfering amplitudes is also discussed.

An important aside is to consider the language used in the literature today for discrete symmetry violation tests.
In the past some of the literature has referred to one of these $CP$ violating triple product asymmetries as a manifestation
of $T$ violation.  Consider the process $\overline{abcd}\to \overline{M}$; this never happens hence it follows that
it is not possible to construct the asymmetry of Eq.~\ref{eq:asym}.  One can invoke \CPT as an argument to call a particular
\CP violation effect a \T violation effect; however that is somewhat futile as it trivialises the problem to the level where
all \CP violation effects are equivalently \T violation effects.  For a long time people have been proposing (and performing) valid
tests of \T violation and searching for \CPT violation.  In particular \T violation has been studied in 
kaons for several decades, culminating in the measurement of Kabir's asymmetry by CPLEAR~\cite{Kabir:1970ts,CPLEAR}.  More recently
Banuls and Bernabeu~\cite{Banuls:2000ki} extended Kabir's approach to use flavour and \CP filter pairs to distinguish between
discrete symmetries for entangled pairs of neutral mesons.  This provides four additional tests of \T, which unlike
Kabir's asymmetry are only tests of \T\footnote{It should be noted that one has to be careful with operator definitions when
allowing \CPT to be violated,
and there is a second possible interpretation that could indicate that the non-zero \T violation measurement 
from \babar could be a \CPT violation
manifest as a fake \T violation~\cite{Fidecaro:2013gsa}.}.  There are additional experiments focused on testing \CPT via edm measurements
as well as using particle decay, mass differences and so on.  
The remainder of these proceedings refer to triple product asymmetries according to
the symmetry under test and does not consider the trivialisation induced by invoking \CPT as worthy of further note.
We can summarise the preceding discussion with the following mis-appropriated quote from Douglas Adams' Hitchiker's
Guide to the Galaxy 
\emph{`` One of the things Ford Prefect had always found hardest to understand about humans was their 
habit of continually stating and repeating the very very obvious''}, such as triple product
asymmetries are not $T$ violating by themselves. It has been noted that with the extra ingredient of 
entanglement we can use triple products to probe $T$ violation (and hence \CPT)~\cite{Bevan:2015ena}.

\section{Are these asymmetries of any use?}

The twelve asymmetries listed above can be measured in a given decay, which prompts the question, what if anything can we learn from them.  A fairly general model, following for example Valencia~\cite{Valencia:1988it}, could be used where one sums over arbitrary $S$, $P$ and $D$ wave amplitudes to identify which of the asymmetries can be driven by the existence of non-zero weak phase differences even in the presence of non-zero strong phase differences.  However it is sufficient to illustrate the point with just two interfering amplitudes given by
\begin{eqnarray}
A_+ &=& a_1 e^{i (\phi_1 + \delta_{1, +})} +  a_2 e^{i (\phi_2 + \delta_{2, +})},\\
A_- &=& a_1 e^{i (\phi_1 + \delta_{1, -})} +  a_2 e^{i (\phi_2 + \delta_{2, -})},\\
\overline A_+ &=& a_1 e^{i (-\phi_1 + \delta_{1, +})} +  a_2 e^{i (-\phi_2 + \delta_{2, +})},\\
\overline A_- &=& a_1 e^{i (-\phi_1 + \delta_{1, -})} +  a_2 e^{i (-\phi_2 + \delta_{2, -})}.
\end{eqnarray}
Here $a_{1,2}$ are the magnitudes of the interfering amplitudes, $\delta_{1,2}$ and $\phi_{1,2}$ are strong and weak phases,
respectively, and the $\pm$ subscript has its usual meaning.  We 
can substitute these amplitudes into the asymmetry equations above to determine the conditions for a non-zero 
asymmetry.  These are as follows:
\begin{eqnarray}
A_P               \!\!&\propto&\!\! r\sin\Delta\phi (\sin\Delta\delta_- -\sin\Delta\delta_+) + r\cos\Delta\phi (\cos\Delta\delta_+ - \cos\Delta\delta_-)\label{eq:tripleproduct:model:ap}\\
\overline A_P     \!\!&\propto&\!\! r\sin\Delta\phi (\sin\Delta\delta_+ - \sin\Delta\delta_-) + r\cos\Delta\phi (\cos\Delta\delta_+ - \cos\Delta\delta_-)\label{eq:tripleproduct:model:apbar}\\
A^P_C             \!\!&\propto&\!\! [(2r^2\cos\Delta \phi \sin[ \Delta \delta_- - \Delta \delta_+]) + r(1+r^2) (\sin\Delta \delta_- - \sin\Delta \delta_+)]\sin\Delta \phi \label{eq:tripleproduct:model:apc}\\
A^\P_{\CP}        \!\!&\propto&\!\! (\cos\Delta\delta_- - \cos\Delta\delta_+)(r^2 (\cos\Delta\delta_- + \cos\Delta\delta_+) + r(1+r^2)\cos\Delta\phi ) \label{eq:tripleproduct:model:apcp}\\
A_C               \!\!&\propto&\!\! 2r \sin[\Delta \delta_-] \sin[\Delta \phi] \\
\overline A_C     \!\!&\propto&\!\! 2r \sin[\Delta \delta_+] \sin[\Delta \phi] \\
A^C_P             \!\!&\propto&\!\! r\left[(1+r^2) (\sin\Delta \delta_- - \sin\Delta \delta_+) + 2r \cos\Delta \phi \sin[ \Delta \delta_- - \Delta \delta_+] \right] \sin \Delta \phi\\
A^\C_{\CP}        \!\!&\propto&\!\! r\left[(1+r^2) (\sin\Delta \delta_- + \sin\Delta \delta_+) + 2r \cos\Delta \phi \sin[ \Delta \delta_- + \Delta \delta_+] \right] \sin \Delta \phi\\
A_{\CP}           \!\!&\propto&\!\! r\cos\Delta\phi (\cos\Delta\delta_+ - \cos\Delta\delta_-) + r\sin\Delta\phi (\sin\Delta\delta_+ + \sin\Delta\delta_-)\\
\overline{A}_{\CP}\!\!&\propto&\!\! r\cos\Delta\phi (\cos\Delta\delta_- - \cos\Delta\delta_+) + r\sin\Delta\phi (\sin\Delta\delta_+ + \sin\Delta\delta_-)\\
A^{\CP}_{\C}      \!\!&\propto&\!\! r\left[(1+r^2)(\sin\Delta\delta_- + \sin\Delta\delta_+) + 2r\cos\Delta\phi\sin(\Delta\delta_- + \Delta\delta_+)  \right]\sin\Delta\phi \\
A^{\CP}_{\P}      \!\!&\propto&\!\! r(\cos\Delta\delta_+ -\cos\Delta\delta_-)[r (\cos\Delta\delta_- + \cos\Delta\delta_+) + (1+r^2)\cos\Delta\phi] .
\end{eqnarray}
where $\Delta\phi = \phi_1 - \phi_2$, $\Delta\delta_\pm = \delta_{1,\pm} - \delta_{2,\pm}$ and $r=a_1/a_2$.
One can see from this that six asymmetries can only be non zero for $\sin\Delta\phi \neq 0$.  These are $A^P_C$, $A_C$, $\overline A_C$, $A^C_P$, $A^\C_{\CP}$, and $A^{\CP}_{\C}$.  
The asymmetries $A_C$, $\overline A_C$ have the familiar 
form of a time-integrated (a.k.a. direct) \CP asymmetry.  
The remaining asymmetries can be non zero under more relaxed conditions that include non-zero strong phase differences even if weak phase differences are zero.  The expected result that ${\cal A}_T = A^P_C \propto \sin\Delta\phi$ can be seen in Eq.~(\ref{eq:tripleproduct:model:apc}).  If one were to extend the restrictive labeling used in some of the literature \emph{true vs fake asymmetries} to these quantities then
the six asymmetries $A^P_C$, $A_C$, $\overline A_C$, $A^C_P$, $A^\C_{\CP}$, and $A^{\CP}_{\C}$ would be called true, 
and the others fake.  For this labeling to be placed correctly in context one has to decode the shorthand; true should be read as ``can only be non-zero for a non-zero weak phase difference''; fake should be read as ``can be non zero for non-zero weak and/or strong phase differences''.

\section{Existing measurements in charm}

The following four body final states have been studied in charm decays: $D^0\to K^+K^-\pi^+\pi^-$, $D^\pm\to K^\pm K_S^0\pi^+\pi^-$, and $D_s^\pm\to K^\pm K_S^0\pi^+\pi^-$~\cite{Link:2000aw,delAmoSanchez:2010xj,Lees:2011dx,Aaij:2014qwa}.  Early measurements by FOCUS reported values of $A_P$, $\overline A_P$, and 
$a^\P_\C$ along with the direct \CP asymmetry.  The results obtained were consistent with zero for all asymmetries.  Following on from this \babar were able to establish non-zero values of $A_P$ and $\overline A_P$ for the $D^0$ and $D_s$ mode.  
The $D^0$ result has been confirmed by LHCb. The results obtained are listed in Table~\ref{tbl:results}.
Recently \babar reanalysed their data in the context of the full set of asymmetries noted here~\cite{ckm2014}.
An interesting issue is highlighted by the LHCb work; previous measurements had integrated over the di-meson invariant 
mass distribution due to the lack of sufficient statistics to perform an amplitude analysis.  With the 
advent of LHCb this limitation goes away and they have produced results binned in $KK$ and $\pi\pi$ invariant masses in
anticipation of performing such an amplitude analysis.  However, in addition to the $KK$ and $\pi\pi$ masses it is
also interesting for the experiments to study $K\pi$ mass distributions.  
These distributions could inform us if there are $S$ or $P$ wave $K\pi$ contributions in the data including, but not limited to, the $K^*(892)$.

\begin{table}[!ht]
\caption{Experimental results published for triple product asymmetries in charm decays}\label{tbl:results}
\begin{center}
{\footnotesize
\renewcommand{\arraystretch}{1.2}
\begin{tabular}{lccc} \hline\hline
 Expt.                      & $A_P$ & $\overline A_P$ & $a^\P_\C$ \\ \hline
\multicolumn{4}{c}{$D^0\to K^+K^-\pi^+\pi^-$}\\
 FOCUS  & & & $0.010\pm 0.057\pm 0.037$ \\ 
 \babar & $-0.069\pm 0.007 \pm 0.006$   & $-0.071\pm 0.007\pm 0.004$    & $0.001\pm 0.005\pm 0.004$ \\
 LHCb   & $-0.0718\pm 0.0041\pm 0.0013$ & $-0.0755\pm 0.0041\pm 0.0012$ & $0.0018\pm 0.0029\pm 0.0004$\\ \hline
\multicolumn{4}{c}{$D^\pm\to K^\pm K_S^0\pi^+\pi^-$}\\
 FOCUS  & & & $0.023\pm 0.062\pm 0.022$\\
 \babar & $+0.011\pm 0.014\pm 0.006$ &$+0.035\pm 0.014\pm 0.007$ & $-0.012\pm 0.010\pm 0.005$\\ \hline
\multicolumn{4}{c}{$D_s^\pm\to K^\pm K_S^0\pi^+\pi^-$}\\ 
 FOCUS & & & $-0.036\pm 0.067\pm 0.023$ \\ 
 \babar & $-0.099\pm 0.011\pm 0.008$ &$-0.072\pm 0.011\pm 0.011$ & $-0.014\pm 0.008\pm 0.003$\\ 
\hline\hline
\end{tabular}
}
\end{center}
\end{table}

Several sensitivity studies have been performed for $D\to VV$ and charm baryon decays at BES III.  These are summarised in
the papers by Kang and Li~\cite{Kang:2009iy,Kang:2010td}.  In many modes one expects that a $\tau$-charm factory 
with 20$\mathrm{fb}^{-1}$ of data would be able to reach (sub)percent level measurements for a number of modes. 
LHCb should be able to measure a number of the modes studied in those references.

\section{Other (potential) measurements}

Interest in the use of triple product asymmetries to probe \CP violation has been around since the 1960s.  In
1993 Heiliger and Seghal predicted that the decay $K_L\to \pi^+\pi^-e^+e^-$ would have a large ${\cal O}(14\%)$
effect manifest~\cite{Heiliger:pipiee}.  
This large effect is the result of an interference between four amplitudes from: $K_L\to\pi^+\pi^-\gamma$ photon conversion;
bremsstrahlung from the \CP violating decay $K_L\to\pi^+\pi^-$; a \CP conserving magnetic dipole component;
and finally a short distance component related to $s\overline{d}\to e^+e^-$.
Shortly afterward this prediction was confirmed by KTeV and subsequently NA48~\cite{Abouzaid:2005te,Lai:2003ad}.  Thus
instead of the typical ${\cal O}(10^{-3})$ or ${\cal O}(10^{-6})$ level of \CP violation associated with 
$\varepsilon$ and $\varepsilon^\prime$ in kaons we observe an ${\cal O}(1)$ effect akin to the magnitudes of \CP violation
effect observed in $B$ decays.  In contrast the corresponding asymmetry measured in $K_S\to \pi^+\pi^-e^+e^-$ is found to be
compatible with zero~\cite{Lai:2003ad}.  This logically follows from the fact that the $K_S\to\pi^+\pi^-$ bremsstrahlung term
is \CP conserving along with the other three contributions in the $K_S$ decay. 
Other systems where one can perform triple product asymmetry measurements include 
$B$ decays, the decays of $b$ and $c$ baryons, and bosonic ($Z$, $H$, or associated production of pairs of 
bosons) decays to four body final states.  It may also be possible to learn something about $\tau$ decays, 
where $\tau$ pairs produced at threshold have the advantage that one constrain the laboratory frame to be
the same as centre of mass frame for the decaying lepton.
Such measurements are discussed in~\cite{Bevan:2014nva}.

\section{Summary}

In summary there are twelve triple product asymmetries that can be used to probe $C$, $P$ and $CP$.
Eight of these have recently been introduced.  Of these asymmetries, for a simple interfering amplitude model,
six can only be non-zero if there is a non-zero weak phase difference.  The remaining six asymmetries
can be non-zero even if the weak phase difference is zero.  Thus there are five new triple product asymmetries
that can provide unambiguous tests of weak interactions.  Collectively the six asymmetries 
$A^P_C$, $A_C$, $\overline A_C$, $A^C_P$, $A^\C_{\CP}$, and $A^{\CP}_{\C}$
provide us with the ability to perform unambiguous tests of $C$, $P$ and \CP 
in the search for non zero weak phase differences.

\Acknowledgements
The author would like to thank the conference organisers for the opportunity to present this work and 
follow up on many useful discussions as a result.


\end{document}